\documentclass{iopart}

\usepackage{iopams}
\usepackage[dvips]{graphicx}

\begin{document}

\title[Heavy flavor kinetics at the hadronization transition]
{Heavy flavor kinetics at the hadronization transition}

\author{D. Blaschke$^{a,b}$, G. Burau$^a$, Yu. Kalinovsky$^c$, 
T. Barnes$^{d,e}$}

\address{$^a$ Fachbereich Physik, Universit\"at Rostock, Germany\\
$^b$ Bogolyubov Laboratory for Theoretical Physics, JINR Dubna, Russia\\
$^c$ Laboratory for Information Technologies, JINR Dubna, Russia\\
$^d$ Oak Ridge National Laboratory, Oak Ridge (TN), USA\\
$^e$ Department of Physics, University of Tennessee, Knoxville TN,
USA}

\ead{david.blaschke@physik.uni-rostock.de}

\begin{abstract}
We investigate the in-medium modification of the charmonium breakup processes
due to the Mott effect for light (pi, rho) and open-charm (D, D*)
quark-antiquark bound states at the chiral/deconfinement phase transition.
The Mott effect for the D-mesons effectively reduces the threshold for
charmonium breakup cross sections, which is suggested as an
explanation of the anomalous $J/psi$ suppression phenomenon in the NA50
experiment.
Further implications of finite-temperature mesonic correlations for the
hadronization of heavy flavors in heavy-ion collisions are discussed.
\end{abstract}


The $J/\psi$ meson plays a key role in the experimental search for the 
quark-gluon plasma (QGP) in heavy-ion collision experiments 
\cite{evidence} where an 
anomalous suppression of its production cross section relative to the 
Drell-Yan continuum as a function of the centrality of the collision has 
been found by the CERN-NA50 collaboration \cite{anomalous}. 
An effect like this has been predicted to signal QGP formation \cite{ms86} 
as a consequence of the screening of color charges in a plasma in close 
analogy to the Mott effect (metal-insulator transition) in dense electronic 
systems \cite{rr}. 
However, a necessary condition to explain $J/\psi$ suppression in the static 
screening model is that a sufficiently large fraction of $c\bar c$ pairs after 
their creation have to traverse regions of QGP where the temperature 
(resp. parton density) has to exceed the Mott temperature 
$T^{\rm Mott}_{{\rm J}/\psi}\sim 1.2 - 1.3~T_c$ \cite{kms,rbs} for a 
sufficiently long time interval $\tau>\tau_{\rm f}$,  where 
$T_c\sim 170$ MeV is the critical phase transition temperature and 
$\tau_{\rm f}\sim 0.3 $ fm/c is the $J/\psi$ formation time. 
Within an alternative scenario \cite{Blaschke:va}, 
$J/\psi$ suppression does not 
require temperatures well above the deconfinement one but can occur already 
at $T_c$ due to impact collisions by quarks from the thermal medium. 
An important ingredient for this scenario is the lowering of the reaction 
threshold for string-flip processes which lead to open-charm meson formation 
and thus to $J/\psi$ suppression. 
This process has an analogue in the hadronic world, where e.g. 
$J/\psi + \pi \rightarrow D^* + \bar D + h.c.$ could occur provided the 
reaction threshold of $\Delta E \sim 640$ MeV can be overcome by pion impact. 
It has been shown recently \cite{Blaschke:2000er,Burau:2000pn} that this 
process and its 
in-medium 
modification can play a key role in the understanding of anomalous $J/\psi$ 
suppression as a deconfinement signal. 
Since at the deconfinement transition the $D$- mesons enter the continuum of 
unbound (but strongly correlated) quark- antiquark states (Mott- effect), the 
relevant threshold for charmonium breakup is lowered and the reaction rate for 
the process gets critically enhanced. Thus a process which is negligible in 
the vacuum may give rise to additional (anomalous) $J/\psi$ suppression when 
conditions of the chiral/ deconfinement transition and $D$- meson Mott effect 
are reached in a heavy-ion collision but the dissociation of the $J/\psi$ 
itself still needs impact to overcome the threshold which is still present but 
dramatically reduced. 
 
For this alternative scenario as outlined in \cite{Burau:2000pn} to work the 
$J/\psi$ 
breakup cross section by pion impact is required and its dependence on the 
masses of the final state $D$- mesons has to be calculated. 
Both, nonrelativistic potential models \cite{Martins:1994hd,wsb00} and chiral 
Lagrangian models \cite{mm98,lk00,hg01} have been employed to determine the 
cross section in the vacuum. The results of the latter models appear to be 
strongly dependent on the choice of formfactors for the meson-meson vertices. 
This is considered as a basic flaw of these approaches which could only be 
overcome when a more fundamental approach, e.g. from a quark model, can 
determine these input quantities of the chiral Lagrangian approach. 
Work in this direction is in progress, see \cite{Blaschke:2000zm}, but 
meanwhile we have to rely on phenomenological approaches.
\begin{figure}[h]
\begin{center}
\includegraphics[height=8cm]{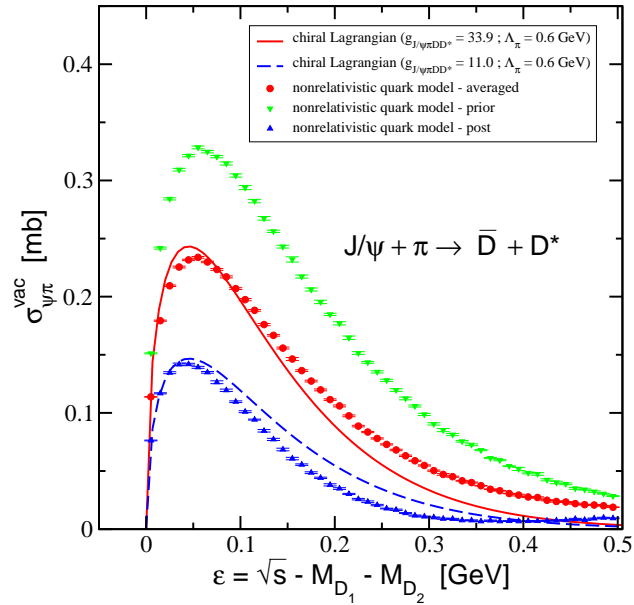}
\caption{Energy dependent breakup cross section for the subprocess
J/$\psi + \pi \to D + \bar D^*$ in the nonrelativistic quark exchange model
\cite{wsb00} (symbols with error bars) and in the chiral Lagrangian model 
\cite{Ivanov:2000vx} (solid and dashed lines).   
 \label{jpsipics} }
\end{center}
\end{figure}
Recently, a modified scheme for defining formfactors of meson-meson vertices 
has been suggested \cite{Ivanov:2000vx} which uses Gaussian functions instead 
of monopole ones \cite{lk00,hg01} and respects differences in the effective
ranges of contact and meson exchange diagrams which arise from their quark
substructure. The result is shown as the solid line in Fig. \ref{jpsipics} and 
compares well with that of the quark exchange model (\cite{wsb00}, filled 
circles) in the peak value and position. The fast drop of the cross section 
above the threshold due to the behavior of the meson wave functions is
described more appropriately by Gaussian formfactors than by monopole ones,
for details see \cite{Ivanov:2000vx}. A change in the coupling constant of
the contact term (which corresponds to the capture diagrams in the quark 
exchange approach) by a factor three leads only to a change in the total cross 
section by about 50 $\%$ (dashed line in Fig. \ref{jpsipics}). 
Thus the dominant contributions come from the meson exchange diagrams, which
are related to the transfer diagrams of Ref. \cite{wsb00}.
Chiral Lagrangian and quark exchange approaches to the hadronic 
interactions of charmonium can be brought to good agreement with each other.

\begin{figure}[h]
\begin{center}
\includegraphics[height=8.5cm]{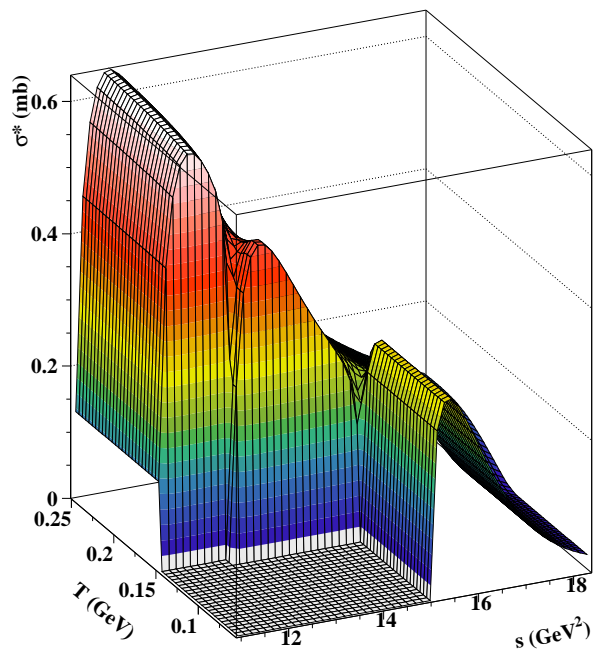}
\caption{In-medium modification of the J/$\psi$ breakup cross section by 
pion impact due to the Mott effect for open charm mesons at $T=150$ MeV.
\label{barnes165} }
\end{center}
\end{figure}

Since at the hadronization transition the $J/\psi$ is expected to be a true 
bound state with almost unchanged mass the most dramatic in-medium modification
for $J/\psi$ breakup processes is expected due to the Mott-effect for pions,
rho- and D- mesons, i.e. their transition from bound states to resonant 
scattering states in the quark-antiquark continuum 
\cite{Blaschke:2000er,Burau:2000pn}. Formally, this transition can be 
described by the behavior of the spectral function in corresponding channel.
As a result, the kinematics for  processes involving heavy mesons is not 
constrained to their (vacuum) mass shell so that they become 
{\it subthreshold}.
The result of a model calculation which uses an off-shell extrapolation of
the quark exchange cross section \cite{wsb00} is given in Fig. \ref{barnes165}.
The corresponding reaction rate for $J/\psi$ dissociation shows a step
function behavior at the Mott (hadronization) transition \cite{Burau:2000pn} 
which has earlier been postulated \cite{Blaizot:1996nq} without a microscopic 
foundation for a successful description of the anomalous $J/\psi$ suppression 
pattern. 
The present approach is a relativistic formulation of string-flip processes
which have been suggested for a unified description of $J/\psi$ 
suppression in hadronic as well as in quark matter \cite{Blaschke:va}.
It predicts a simultaneus increase in the reaction rates for breakup of all 
hidden charm (bottom) states once the Mott effect for the final state 
open charm (bottom) hadrons occurs. Upon neglect of this trigger mechanism 
one would predict instead a sequential breakup of heavy quarkonia according to
their different dissociation energies into the quark-antiquark continuum 
\cite{Nardi:1998qb,Digal:2001ue,Wong:2001an}.

Applications to the description of charmonium production in nucleus-nucleus
collisions are in progress \cite{bbp}.

In summary, we have shown that recent developments of the chiral Lagrangian 
and quark exchange model appraches have lead to a convergence of results for 
quarkonium-hadron breakup (string-flip) cross sections. Reaction rates for
these processes are negligibly small in the hadronic world but show a critical 
enhancement at the deconfinement/ hadronization transition when the Mott 
effect for open flavor hadrons is taken into account. This reduces 
dramatically those reaction thresholds which would otherwise
prevent heavy flavor equilibration in hadronic matter.
The inclusion of the Mott effect into the description of heavy flavor kinetics 
at the hadronization transition allows to link three recently observed 
phenomena: 
(i) anomalous $J/\psi$ suppression \cite{anomalous,Blaschke:2000er,bbp}, 
(ii) open charm enhancement \cite{NA50open,hf_enh},
and (iii) the success of the statistical model 
\cite{Br1,Ko:01,Grandchamp:2001pf}.   

\ack

The research of Yu.K. has been supported by DFG grant No. 436 RUS
17/62/01. D.B. and G.B. acknowledge support by DAAD for their
visits at Oak Ridge National Laboratory and University of Tennessee 
at Knoxville. T.B. has been supported in part by NSF grant No. INT-0004089.

\end{document}